\documentclass[journal=jacsat,manuscript=article]{achemso}

\usepackage{chemformula} % Formula subscripts using \ch{}
\usepackage[T1]{fontenc} % Use modern font encodings

\usepackage{graphicx}
\usepackage{color}      % use if color is used in text
\usepackage[english]{babel}
\usepackage{amsmath,amsfonts,amssymb}

\addto\captionsenglish{}

\author {Emilie Sakat}
\affiliation[LCF]{Universite Paris-Saclay, Institut d'Optique Graduate School, CNRS, Laboratoire Charles Fabry, 91127, Palaiseau, France}
\alsoaffiliation[C2N]{Universite Paris Saclay, Center for Nanoscience and Nanotechnology, C2N UMR 9001, CNRS, 91120 Palaiseau, France}
\email{emilie.sakat@c2n.upsaclay.fr}
\author {L\'eo Wojszvzyk}
\affiliation[LCF]{Universite Paris-Saclay, Institut d'Optique Graduate School, CNRS, Laboratoire Charles Fabry, 91127, Palaiseau, France}
\author {Jean-Jacques Greffet}
\affiliation[LCF]{Universite Paris-Saclay, Institut d'Optique Graduate School, CNRS, Laboratoire Charles Fabry, 91127, Palaiseau, France}
\author {Jean-Paul Hugonin}
\affiliation[LCF]{Universite Paris-Saclay, Institut d'Optique Graduate School, CNRS, Laboratoire Charles Fabry, 91127, Palaiseau, France}
\author {Christophe Sauvan}
\affiliation[LCF]{Universite Paris-Saclay, Institut d'Optique Graduate School, CNRS, Laboratoire Charles Fabry, 91127, Palaiseau, France}
\email{christophe.sauvan@institutoptique.fr}

\title[An \textsf{achemso} demo]
  {Upper bound for light absorption assisted by a nanoantenna}

\abbreviations{IR,NMR,UV}

\begin{document}

%\begin{tocentry}
%
%Some journals require a graphical entry for the Table of Contents.
%This should be laid out ``print ready'' so that the sizing of the
%Inside the \texttt{tocentry} environment, the font used is Helvetica
%8\,pt, as required by \emph{Journal of the American Chemical
%The surrounding frame is 9\,cm by 3.5\,cm, which is the maximum
%permitted for  \emph{Journal of the American Chemical Society}
%graphical table of content entries. The box will not resize if the
%content is too big: instead it will overflow the edge of the box.
%
%This box and the associated title will always be printed on a
%separate page at the end of the document.
%
%\end{tocentry}

\begin{abstract}
  We study light absorption by a dipolar absorber in a given environment, which can be a nanoantenna or any complex inhomogeneous medium. From first-principle calculations, we derive an upper bound for the absorption, which decouples the impact of the environment from the one of the absorber. Since it is an intrinsic characteristic of the environment regardless of the absorber, it provides a good figure of merit to compare the ability of different systems to enhance absorption. We show that, in the scalar approximation, the relevant parameter is not the field enhancement but the ratio between the field enhancement and the local density of states. Consequently, a plasmonic structure supporting hot spots is not necessarily the best choice to enhance absorption. We also show that our theoretical results can be applied beyond the scalar approximation and the plane-wave illumination.
\end{abstract}

\section{Introduction}
Light absorption and emission are two fundamental processes of light-matter interaction~\cite{Beer,Einstein}. Absorption converts electromagnetic energy carried by photons to internal energy of matter carried by electrons or phonons. Controlling the absorption in small volumes is a major issue for numerous applications, such as photovoltaics~\cite{ReviewAtwaterPolman,Vynck2012,Mubeen2011}, sensing by surface-enhanced infrared absorption (SEIRA)~\cite{Baldassarre2015,Giessen2017,Dong2017} or enhanced infrared photoexpansion nanospectroscopy~\cite{Jin2019}, thermal emission~\cite{Sakat2018,Greffet2018}, photothermal effects at the nanoscale for enhanced photocatalysis and nano-chemistry~\cite{Christopher2010,Carlson2012,Baffou2013}.

Absorption inside subwavelength objects is weak but different strategies can be used to circumvent this limitation. The absorber properties, the illumination, or the environment can be engineered. Many works have been dedicated to the optimization of the absorption by a single nanoparticle \cite{Tretyakov2014,Fleury2014,Miller2014,Grigoriev2015,Colom2016,Miller2016,Ivanenko2019}. For an absorber in a homogeneous medium, the physical bounds of the problem are known; they depend on the multipolar character of the particle. The maximum absorption cross-section of a molecule or a nanoparticle that behaves like a pure electric dipole (dipolar approximation), is $3\lambda^2/(8\pi n^2)$, with $\lambda$ the wavelength and $n$ the surrounding refractive index~\cite{Tretyakov2014,Grigoriev2015}. This upper bound can only be reached if the absorber polarizability $\alpha$ matches a precise value, $\alpha_\mathrm{opt} = i3\lambda^3/(8\pi^2 n)$. Subwavelength particles that go beyond the dipolar approximation offer additional degrees of freedom and are governed by different physical bounds~\cite{Fleury2014,Miller2014,Grigoriev2015,Colom2016,Miller2016,Ivanenko2019}. Absorption by an ensemble of particles in a homogeneous medium is yet another related problem with a different upper limit~\cite{Hugonin2015}.

\begin{figure}[b]
	\centerline{\includegraphics[width=0.6\columnwidth]{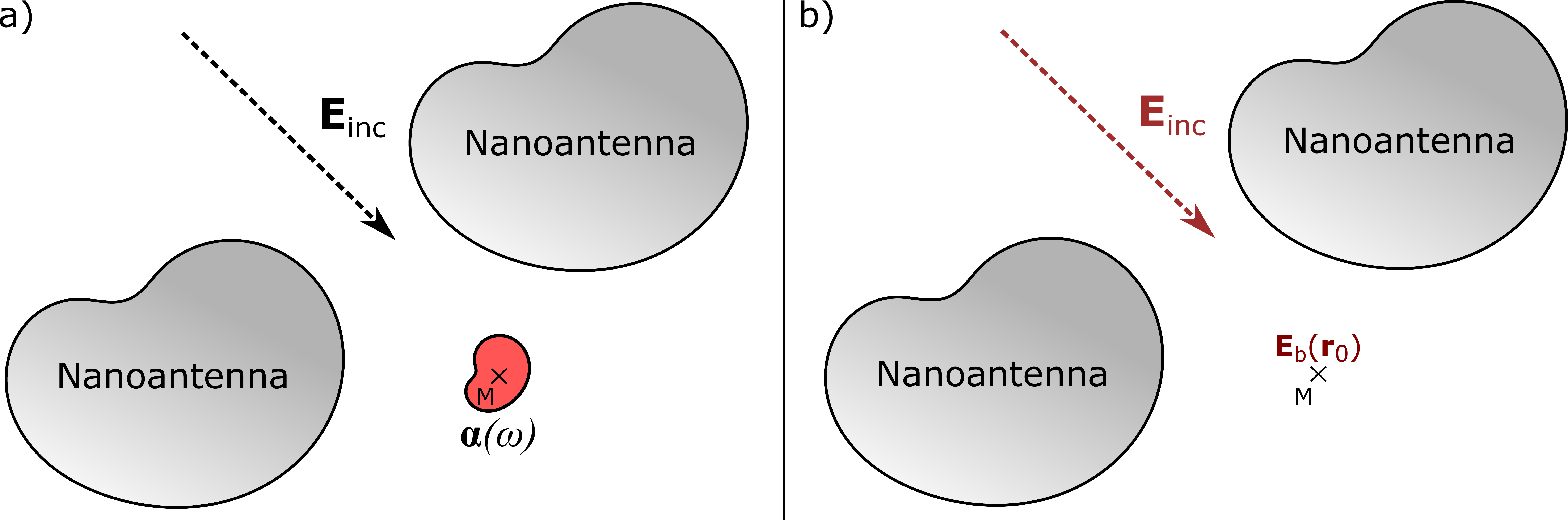}}
	\caption{(a) Subwavelength absorber (in red) in an arbitrary environment (in gray, referred to as "nanoantenna") illuminated by an incident field ${\bf E}_\mathrm{inc}$. We focus on the absorption inside the particle. (b) Same problem without the absorber. The field ${\bf E}_b$ is the field scattered by the environment alone.}
	\label{fig:principe}
\end{figure}

Plunging the absorber in a complex medium or \mbox{modifying} the illumination offers additional possibilities to tailor the absorption~\cite{Sakat2018,Castanie2012,Sentenac2013}. In this work, we consider a subwavelength absorber in a complex environment (inhomogeneous or not) as depicted in Fig.~\ref{fig:principe}. We focus on the absorption inside the absorber and do not discuss dissipation in the surroundings, if any. To fully exploit the control possibilities offered by the environment, it is crucial to know what is the relevant figure of merit. The absorption density is proportional to the local electric-field intensity $|{\bf E}({\bf r})|^2$~\cite{Jackson}, which results from the environment \emph{and} the absorber. It is highly desirable to decouple both contributions to provide a figure of merit that is intrinsic to the environment. Only then can we properly compare the ability of different structures to modify absorption.

Let us draw a parallel with spontaneous emission. In a complex medium, it is modified by a factor proportional to the photonic local density of states (LDOS) \cite{NovotnyBook}. For an emitter coupled to a resonant cavity, the emission enhancement has a well-known upper bound, the Purcell factor, which is intrinsic to the resonator and independent of the emitter~\cite{Purcell,Haroche,Sauvan2013}. The upper bound is reached if emitter and resonator fulfill a few matching conditions --~spectral, spatial, and in polarization~\cite{Sauvan2013}.

The link between the absorption and the environment properties is more complex than the proportionality relation between emission and LDOS. Let us start by the simple case where the absorber perturbs only slightly the electromagnetic field. In this perturbation regime, the Born approximation can be used and the field inside the absorber is asssumed to be equal to the field ${\bf E}_b$ provided by the environment alone, see Fig.~\ref{fig:principe}b. In that case, the absorption is simply proportional to the field enhancement $|{\bf E}_b|^2/|{\bf E}_\mathrm{inc}|^2$. Hence, hot spots are often thought to provide large absorption enhancements and nanoantennas are often engineered for maximizing the field enhancement they provide~\cite{Seok2011,Metzger2014}. However, if the presence of the absorber significantly affects the local field, a perturbative treatment is not valid and the situation is more complex. It has been recently shown that both the field enhancement and the LDOS play an intricate role in the absorption~\cite{Castanie2012}. However, no clear upper bound --~kind of Purcell factor analogue~-- has been derived for the general problem of an absorber in a complex environment.

Antenna theory provides a solution in one specific case. For an antenna receiving a signal from the direction $(\theta,\phi)$, the maximum absorption cross-section of the load is $G_a(\theta,\phi)\lambda^2/(4\pi n^2)$, with $G_a$ the antenna gain~\cite{Balanis,Andersen2005,Radi2013}. This upper bound is reached if the load is impedance-matched with the antenna. The gain is defined for an emitting antenna as the fraction of the wall-plug power radiated in the direction $(\theta,\phi)$. Unfortunately, the $G_a\lambda^2/(4\pi n^2)$ limit only applies to plane-wave illumination and to antennas working in the scalar approximation when one component of the electromagnetic field is dominant. In cases where the vectorial nature of the electromagnetic field cannot be neglected, the problem remains open. Moreover, the antenna point of view highlights the gain, whereas other derivations underline the field enhancement and the LDOS~\cite{Castanie2012}. It is thus important to generalize existing results, while enlightening the link between them.

In this work, we derive a general upper bound for the power dissipated in a subwavelength absorber surrounded by any complex inhomogeneous environment. Owing to its subwavelength dimensions, the absorber is treated in the electric-dipole approximation. On the other hand, the electromagnetic properties of the environment are treated rigorously without any approximation. The upper bound is independent of the absorber; it entirely depends on the environment and the illumination. Thus, it provides a relevant figure of merit for comparing the ability of different systems to enhance absorption. This figure of merit results from the interplay between two electromagnetic properties of the environment, the field enhancement and the Green tensor. We show under which assumptions the gain is a relevant parameter. We finally discuss under which conditions the system can reach the upper bound. We apply the theory to a few emblematic examples of nanophotonics: a plasmonic dimer nanoresonator, dielectric nanoantennas, and a silicon-on-insulator (SOI) ridge waveguide. We evidence that a plasmonic system providing extremely large field enhancements is not necessarily an optimal choice to increase absorption.

\section{Results and discussion}

We consider the problem illustrated in Fig.~\ref{fig:principe}(a). An absorbing subwavelength particle (in red) is placed in a complex absorbing or non-absorbing environment (in gray) and illuminated by an incident field ${\bf E}_\mathrm{inc}$. We focus on the absorption inside the particle. We consider passive materials and use the $\exp(-i\omega t)$ convention for time-harmonic fields, with $\omega = 2\pi c/\lambda$. The following derivations apply to any incident field and any environment geometry. We refer to the environment as ``the antenna'', but no particular assumption is made on its geometry.

The total electric field can be written as ${\bf E}({\bf r}) \equiv {\bf E}_b({\bf r}) + {\bf E}_s({\bf r})$, where ${\bf E}_b$ is the field \emph{in the absence of} the absorber, see Fig.~\ref{fig:principe}(b), and ${\bf E}_s$ is the field scattered by the absorber. It is the field radiated by the current source induced inside the absorber by the exciting field. We now make the sole assumption of our derivation. We assume that the subwavelength particle scatters light like an electric dipole ${\bf p}$ located at ${\bf r} = {\bf r}_0$. The scattered field is then ${\bf E}_s({\bf r}) = \mu_0 \omega^2 {\bf G}({\bf r},{\bf r}_0){\bf p}$, with ${\bf G}$ the Green tensor of the antenna alone.

The power dissipated in the particle is the difference between extinction and scattering, $P_a = P_e \!-\! P_s$~\cite{Jackson}. Within the dipole approximation, $P_e = - \frac{1}{2}\omega \mathrm{Im}({\bf p}^\dagger {\bf E}_b)$ and $P_s = \omega {\bf p}^\dagger {\bf g}{\bf p}$, with ${\bf g} \equiv \frac{1}{2}\mu_0\omega^2 \mathrm{Im}[{\bf G}({\bf r}_0,{\bf r}_0)]$ [SI]. It follows

\begin{equation}\label{eq:Pabs}
    P_a = - \frac{1}{2}\omega \mathrm{Im}({\bf p}^\dagger {\bf E}_b) - \omega {\bf p}^\dagger {\bf g}{\bf p}  \,,
\end{equation}

\noindent where ${\bf E}_b \!\equiv\! {\bf E}_b({\bf r}_0)$ and ${\bf p}^\dagger$ is the conjugate transpose of ${\bf p}$. The induced dipole is ${\bf p} = \varepsilon_0 \boldsymbol{\alpha}(\omega) [\mathbf{E}_b({\bf r}_0) + \mu_0 \omega^2\mathbf{S}({\bf r}_0,{\bf r}_0) \mathbf{p} ]$, with $\boldsymbol{\alpha}(\omega)$ the polarizability tensor of the particle and $\mathbf{S} = \mathbf{G} - \mathbf{G}_0$, $\mathbf{G}_0$ being the Green tensor of the homogeneous medium of refractive index $n$ that surrounds the particle.

The dissipated power can be rewritten as [SI]

\begin{equation}\label{eq:Inegalite}
    P_a = -\omega \Big(\mathbf{p}^\dagger + \frac{i}{4}\mathbf{E}_b^\dagger \mathbf{g}^{-1}\Big) \mathbf{g} \Big(\mathbf{p} - \frac{i}{4}\mathbf{g}^{-1}\mathbf{E}_b \Big) + \frac{\omega}{16}\mathbf{E}_b^\dagger\mathbf{g}^{-1}\mathbf{E}_b .
\end{equation}

\noindent Since ${\bf g}$ is a positive semi-definite matrix~\footnote{In a complex medium with only passive materials, whatever the complex vector ${\bf u}$, the quantity $\omega {\bf u}^\dagger {\bf g}{\bf u}$ is the power emitted by the dipole source ${\bf u}$. Thus ${\bf u}^\dagger {\bf g}{\bf u} \geqslant 0$.}, we know that $-\omega (\mathbf{p}^\dagger + \frac{i}{4}\mathbf{E}_b^\dagger \mathbf{g}^{-1}) \mathbf{g} (\mathbf{p} - \frac{i}{4}\mathbf{g}^{-1}\mathbf{E}_b ) \leqslant 0$. This readily leads to an upper bound for the absorption, $P_a \leqslant P_a^\mathrm{max}$,

\begin{equation}\label{eq:EqPmax}
    P_a^\mathrm{max} = \frac{\omega}{16}\mathbf{E}_b^\dagger\mathbf{g}^{-1}\mathbf{E}_b \,.
\end{equation}

\noindent According to Eq.~\eqref{eq:Inegalite}, the upper bound is reached for an optimal dipole ${\bf p}_\mathrm{opt} = \frac{i}{4}\mathbf{g}^{-1}\mathbf{E}_b$. The optimal polarizability that yields the maximum absorption is then [SI]

\begin{equation}\label{eq:EqM2}
    \boldsymbol{\alpha}_\mathrm{opt}(\omega) = \frac{c^2}{\omega^2}\left [ \mathbf{S}^*({\bf r}_0,{\bf r}_0)- i\frac{\omega n}{3\pi c} \mathbf{I} \right ]^{-1} .
\end{equation}

\noindent with $\mathbf{S}^*$ the conjugate of $\mathbf{S}$. Any other polarizability in the same environment necessarily absorbs less light. Equation~\eqref{eq:EqM2} can be seen as a vectorial generalization of the usual scalar impedance-matching concept~\cite{Balanis,Greffet2010}.

We define the absorption efficiency $\eta_a = P_a/P_\mathrm{inc}$ as the fraction of the incident power that is absorbed inside the particle. The maximum absorption efficiency is

\begin{equation}\label{eq:EqEffmax}
    \eta_a^\mathrm{max} = \frac{\omega}{16}\frac{\mathbf{E}_b^\dagger\mathbf{g}^{-1}\mathbf{E}_b}{P_\mathrm{inc}} .
\end{equation}

\noindent If the incident field is a plane wave, we rather define a maximum absorption cross-section by normalizing $P_a^\mathrm{max}$ with the incident Poynting vector $\frac{1}{2}\varepsilon_0 cn|{\bf E}_\mathrm{inc}|^2$,

\begin{equation}\label{eq:EqM1}
    \sigma_a^\mathrm{max} = \frac{3\lambda^2}{8\pi n^2} g_0 \frac{\mathbf{E}_b^\dagger \mathbf{g}^{-1} \mathbf{E}_b}{|{\bf E}_\mathrm{inc}|^2} ,
\end{equation}

\noindent where $g_0\mathbf{I} \!\equiv\! \frac{1}{2}\mu_0\omega^2 \mathrm{Im}[{\bf G}_0({\bf r}_0,{\bf r}_0)] \!=\! \omega^3n/(12\pi\varepsilon_0 c^3) \mathbf{I}$. Note that, in a homogeneous environment, ${\bf g} \!=\! g_0\mathbf{I}$, $\mathbf{E}_b \!=\! {\bf E}_\mathrm{inc}$, and we recover the usual result $\sigma_a^\mathrm{max} = 3\lambda^2/(8\pi n^2)$.

Equations~\eqref{eq:EqPmax}-\eqref{eq:EqM1} form the central result of this article. They deserve a few comments before we illustrate their consequences on a couple of examples. The upper bound is independent of the absorber; it solely depends on the antenna and the incident field. Thus, Eqs.~\eqref{eq:EqEffmax} and~\eqref{eq:EqM1} provide meaningful figures of merit for comparing the ability of different systems to enhance absorption. These novel figures of merit result from an interplay between the local field $\mathbf{E}_b$ provided by the sole antenna and the imaginary part of its Green tensor $\mathbf{g}$.

In the case where the vectorial nature of the electromagnetic field can be neglected, the general upper bound can be replaced by an approximate form. Let us assume that the field $\mathbf{E}_b$ and the Green tensor $\mathbf{g}$ are dominated by a single component, the $z$ component. Within this scalar approximation, Eq.~\eqref{eq:EqM1} reduces to

\begin{equation}\label{eq:EqM1scalar}
    \sigma_a^\mathrm{max} \approx \frac{3\lambda^2}{8\pi n^2} \frac{|E_{bz}|^2}{|E_\mathrm{inc}|^2} \frac{g_0}{g_{zz}} = \frac{3\lambda^2}{8\pi n^2} \frac{\mathrm{Intensity \,\, enh.}}{\mathrm{LDOS \,\, enh.}} \,.
\end{equation}

\noindent The upper bound appears to be the ratio between the intensity enhancement $|E_{bz}|^2/|E_\mathrm{inc}|^2$ and the LDOS enhancement $g_{zz}/g_0$. The fact that a LDOS enhancement is detrimental can be intuitively understood as follows: a larger LDOS increases the radiation of the induced dipole, i.e., it increases the scattering at the expense of absorption.

It can be shown using reciprocity that Eq.~\eqref{eq:EqM1scalar} is equivalent to $\sigma_a^\mathrm{max} = G_a(\theta,\phi)\lambda^2/(4\pi n^2)$, with $G_a$ the antenna gain [SI]. The upper bound derived with antenna theory is thus a particular (scalar) case of the general result in Eq.~\eqref{eq:EqM1}. Two equivalent points of view can be adopted. They provide different conclusions and it is interesting to consider both of them. Equation~\eqref{eq:EqM1scalar} evidences that antennas supporting hot spots are a good choice if, \emph{and only if}, the formation of hot spots is not concomitant with a large LDOS. On the other hand, the antenna gain is the product of the radiative efficiency and the directivity, $G_a(\theta,\phi) = \eta_r D(\theta,\phi)$~\cite{Balanis}. This second point of view underlines that (i) directional structures perform better and (ii) dielectric objects ($\eta_r = 1$) are better choices than plasmonic structures with $\eta_r < 1$.

Let us illustrate the theory with a few examples. First, we study a plasmonic dimer nanoresonator and a Yagi-Uda antenna made of silicon nanospheres. Both can be treated in the scalar approximation. Secondly, we consider structures that can only be studied with a fully vectorial formalism: silicon nanospheres in a L-shape configuration and a SOI ridge waveguide.

\begin{figure*}
    \includegraphics[width=\columnwidth]{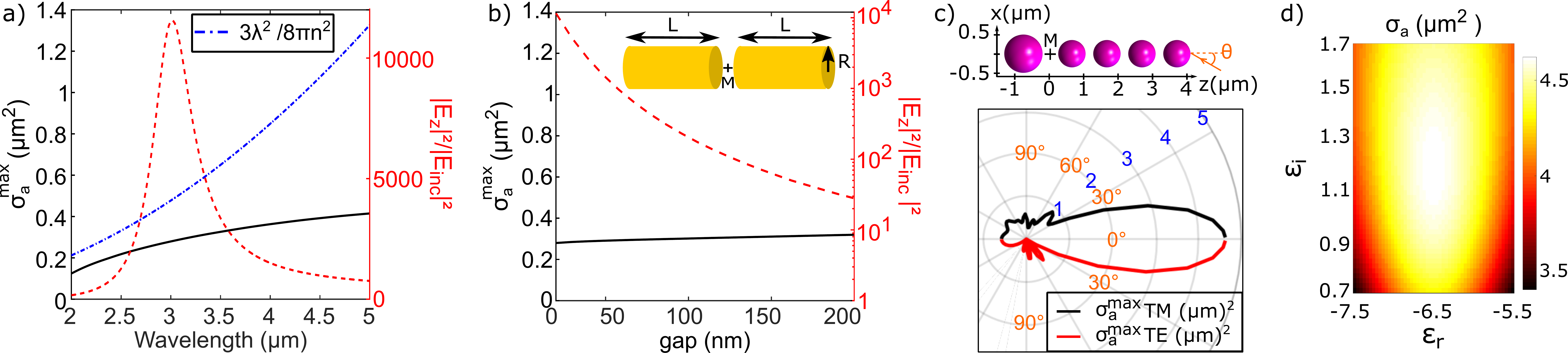}
	\caption{Nanoantennas in the scalar approximation. (a)-(b) Dimer composed of two gold cylinders illuminated by a normal-incident plane wave polarized along the cylinder axis. The point $M({\bf r}_0)$ is the gap center. (a) Spectra of the maximum absorption cross-section $\sigma_a^\mathrm{max}$ (left axis, black curve) and the intensity enhancement (right axis, dashed red curve) for $L \!=\! 414$~nm, $R \!= \!15$~nm, and $w \!=\! 20$~nm. (b) Maximum absorption cross-section and intensity enhancement as a function of the gap width $w$ (the cylinder length is tuned to keep the resonance fixed at $\lambda=3~\mu$m). (c)-(d) Yagi-Uda antenna composed of silicon spheres (refractive index 3.436). The reflector (radius 498~nm) is separated from the first director by 700~nm. The four directors (radius 382~nm) are equally spaced by 250~nm. The point $M({\bf r}_0)$ is the center between the reflector and the first director. (c) Maximum absorption cross-section $\sigma_a^\mathrm{max}$ at $\lambda=3~\mu$m as a function of the incident angle $\theta$ for a TM (black curve) or TE (red curve) polarized plane wave. (d) Absorption cross-section of an absorbing nanosphere of radius 176~nm inserted at ${\bf r}_0$. The real and imaginary parts of the absorber permittivity are varied around the optimal value $\varepsilon_\mathrm{opt}$=\mbox{-6.46+1.22i}. (a)-(b) Calculations performed with an aperiodic Fourier modal method dedicated to body-of-revolution objects~\cite{Bigourdan2014}. The gold permittivity is given in [SI]. (c)-(d)  Calculations performed with a multipole method~\cite{Stout2002,Stout2011}. In all cases, the antenna is embedded in a medium of refractive index $n=1.5$.}
	\label{fig:dimer}
\end{figure*}

In the case of a single-mode antenna, LDOS and field enhancement are not independent since both are driven by the excitation of an eigenmode~\cite{ReviewQNM}. Both are resonant with the same spectral profile and increasing one necessarily enhances the other. Thus, quite counter-intuitively, the value of the upper bound in Eq.~\eqref{eq:EqM1scalar} \emph{is not correlated to the presence of hot spots}. To illustrate this conclusion, let us consider a dimer made of two gold nanorods illuminated by a plane wave. Only the dipole mode at $\lambda \!=\! 3~\mu$m is excited. Figure~\ref{fig:dimer}(a) displays the spectra of the maximum absorption cross-section and the intensity enhancement at the gap center: $|{\bf E}_b|^2/|{\bf E}_\mathrm{inc}|^2$ is resonant whereas $\sigma_a^\mathrm{max}$ is not. Figure~\ref{fig:dimer}(b) shows the variation of $|{\bf E}_b|^2/|{\bf E}_\mathrm{inc}|^2$ and $\sigma_a^\mathrm{max}$ with $w$. As $w$ is varied, the cylinder length is tuned to keep the resonance fixed at $\lambda \!=\! 3~\mu$m so that all the antennas are at resonance. The maximum absorption cross-section only weakly varies while the field enhancement strongly decreases. It is noteworthy that, since $\sigma_a^\mathrm{max} < 3\lambda^2/(8\pi n^2)$, any absorber in the gap of this plasmonic dimer absorbs less than an optimal absorber in a homogeneous medium.

Since the upper bound in the scalar case is proportional to the radiative efficiency $\eta_r$ and the directivity, we switch to a Yagi-Uda antenna made of silicon nanospheres, which is known to be directional \cite{Stout2011}. We keep a plane-wave illumination and a wavelength of $\lambda=3~\mu$m. Thus, silicon is transparent and $\eta_r = 1$. The maximum absorption cross-section $\sigma_a^\mathrm{max}$ is represented on Fig.~\ref{fig:dimer}(c) as a function of the incident angle $\theta$ for ${\bf r}_0$ located in between the reflector and the first director. The incident plane wave is either TM (black curve) or TE (red curve) polarized. For $\theta=0^\circ$, $\sigma_a^\mathrm{max}$ is one order of magnitude larger than $3\lambda^2/(8\pi n^2)$. This evidences that a dielectric directional antenna can provide larger absorption enhancements than a plasmonic antenna supporting hot spots. Note that, for a fixed geometry, the upper bound changes with the incident angle, whereas the optimal polarizability given by Eq.~\eqref{eq:EqM2} is constant. Once the polarizability has been chosen, one is sure to reach the upper bound whatever the incident field.

Before considering fully vectorial examples, we discuss the possibility to reach the upper bound on the Yagi-Uda example. The optimal polarizability corresponds for instance to a sphere of radius 176~nm filled with a material of relative permittivity $\varepsilon_\mathrm{opt} = -6.46+1.22i$~\footnote{We have rigorously calculated the relation between the electric-dipole polarizability and the sphere radius and permittivity with Mie theory}. Such a permittivity at $\lambda \!=\! 3~\mu$m can be obtained with highly-doped semiconductor nanocrystals \cite{Mendelsberg2012,Mendelsberg2012a}. We have checked with a rigorous numerical method that inserting this nanosphere in the Yagi-Uda antenna yields an absorption that is indeed equal to the upper bound [SI]. We have also tested the robustness of the optimal polarizability. Figure~\ref{fig:dimer}(d) shows the absorption cross-section at normal incidence in TM polarization obtained by varying the absorber permittivity around the optimum. Since the cross-section varies smoothly around the maximum, we have a good flexibility on the choice of the permittivity.

\begin{figure}[ht]
    \centerline{\includegraphics[width=0.55\columnwidth]{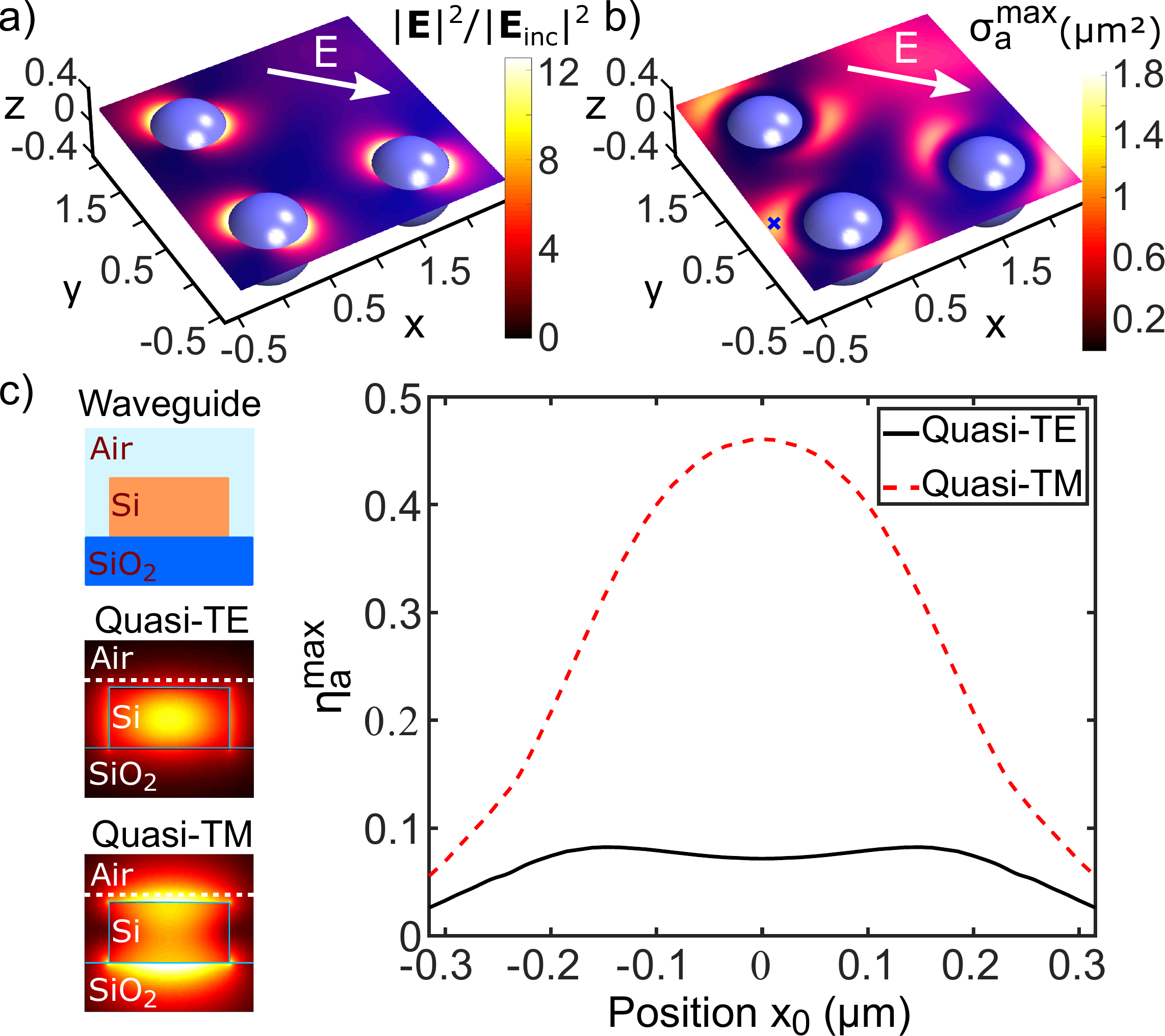}}
	\caption{Beyond scalar approximation. (a)-(b) L-shape antenna made of three silicon spheres (radius 430~nm) equally spaced by $1.54~\mu$m and embedded in a medium of index $n=1.5$ at $\lambda=3~\mu$m. (a) Intensity enhancement and (b) maximum absorption cross-section as a function of the position ${\bf r}_0$ in the $(x,y)$ plane for $z=232$~nm. (c) SOI ridge waveguide at $\lambda=1.5~\mu$m. The silicon ridge ($n=3.5$, width 500~nm, thickness 250~nm) lies over a silica substrate ($n=1.44$). Maximum absorption efficiency as a function of the position along the dashed white line located 20~nm above the waveguide for the fundamental quasi-TE mode (solid black curve) and the quasi-TM mode (dashed red curve). The insets give the electric-field intensity distributions of both modes.}
	\label{fig:AntenneL}
\end{figure}

Let us now consider a L-shape antenna made of three silicon spheres, see Figs.~\ref{fig:AntenneL}(a)-(b). Such a structure cannot be described in the scalar approximation, especially for positions ${\bf r}_0$ outside the symmetry planes. In that case, Eq.~\eqref{eq:EqM1scalar} is not valid. The system can only be characterized by the vectorial upper bound of Eq.~\eqref{eq:EqM1}. The latter depends on the absorber position and it is important, for a given nanoantenna, to evaluate where the particle could reach the best absorption. Figures~\ref{fig:AntenneL}(a)-(b) show respectively the spatial distributions of the intensity enhancement and the maximum absorption cross-section in the $(x,y)$ plane for $z=232$~nm. The antenna is illuminated from the bottom by a plane wave propagating along the $z$ axis and polarized linearly along the white arrow. The blue cross in Fig.~\ref{fig:AntenneL}(b) marks the position where the upper bound is maximum. It does not correspond to the maximum of $|{\bf E}_b|^2/|{\bf E}_\mathrm{inc}|^2$, which evidences, once again, that seeking hot spots is not sufficient to maximize the absorption.

Let us study a last example where the incident field is not a plane wave. It allows us to evidence the generality of the absorption upper bound that can be used in a variety of photonic structures. We consider an absorber located above a SOI ridge waveguide. In this integrated configuration, the incident field is a guided mode. By applying Eq.~\eqref{eq:EqEffmax}, we have calculated the maximum absorption efficiency above the waveguide, see Fig.~\ref{fig:AntenneL}(c). The absorber is moved along the horizontal white line located 20~nm above the waveguide; the incident field is either the fundamental quasi-TE mode (solid black curve) or the quasi-TM mode (dashed red curve). The latter provides an absorption efficiency of almost $50\%$, even for an absorber located outside the waveguide core.

\section{Conclusion}

In conclusion, we derived an upper bound for the problem of absorption by a dipolar absorber in a complex environment, homogeneous or not. The derivation relies on a vectorial formalism and is valid for any environment and any illumination. Since it decouples the environment from the absorber, the upper bound provides a meaningful figure of merit for comparing the intrinsic ability of different structures to enhance absorption in a nanovolume. Moreover, it allows seeking, for one given structure, the optimal position for the absorber. In the scalar approximation, the relevant parameter is not the field enhancement but the ratio between field enhancement and LDOS. Thus, although placing an absorber in a hotspot of a plasmonic structure can increase the absorption~\cite{Sakat2018}, plasmonic antennas supporting hot spots are not necessarily the best candidates to reach high absorption enhancements. Larger enhancements can be achieved with dielectric directional antennas. In cases where the scalar approximation is not valid or the illumination is not a plane wave, the situation is more complex and our general upper bound is the good figure of merit to characterize the problem. Since it relies on electromagnetic calculations (field enhancement and Green tensor) that are nowadays performed routinely, we think that this work opens new avenues for understanding and optimizing absorption in subwavelength volumes inserted in complex media. It should benefit various applications such as SEIRA, photovoltaics, thermal emission or nanochemistry. This work could also serve as a building block for the study of multipolar and/or multiple absorbers.

%\subsection{References}


%\begin{acknowledgement}
%
%Please use ``The authors thank \ldots'' rather than ``The
%authors would like to thank \ldots''.
%
%\end{acknowledgement}

\newpage
% \appendix*
 
%\begin{suppinfo}
\section{Supporting Information}

%Upper bound for light absorption assisted by a nanoantenna}

%\maketitle
%\vspace{0.5cm}

We provide hereafter some additional elements concerning the numerical calculations and the derivations of the upper bound and of the optimal polarizability presented in the main text.

\begin{enumerate}
  \item Section A contains the Drude model that has been used for the gold permittivity.
  \item Section B presents an original demonstration of the power dissipated by a dipolar absorber.
  \item Section C provides more details on the derivation of the upper bound for the absorption.
  \item Section D describes in more details the derivation of the optimal polarizability.
  \item Section E concerns the specific case of scalar approximation and plane-wave illumination. It presents the derivation, based on reciprocity, of the relation between the field enhancement and the antenna gain.
  \item Section F shows the possibility to decompose the vectorial problem in the sum of three scalar problems in the case of reciprocal materials.
  \item Section G presents a numerical validation of our theoretical results. We compare our theoretical predictions with a rigorous calculation of the absorption cross-section of a nanosphere with the optimal polarizability inserted in the Yagi-Uda nanoantenna studied in the main text.

\end{enumerate}

\subsection{A. Drude model of the gold permittivity}

The dielectric permittivity of gold that we have used for the calculations is given by a Drude model that fits the data tabulated in~\cite{Olmon2012} over the $[2-5] \,\mu$m  spectral range,

\begin{equation}
    \varepsilon(\omega) = \varepsilon_{\infty} - \frac{\omega_p^2}{\omega^2 + i\omega\gamma} \,,
\end{equation}

\noindent with $\omega = 2\pi c/\lambda$ the angular frequency, $\varepsilon_{\infty} = -2.1503$ the high-frequency permittivity, $\omega_p = 1.223 \times 10^{16}\,\mathrm{rad.s}^{-1}$ the plasma frequency, and $\gamma = 7.4341 \times 10^{13}\,\mathrm{ rad.s}^{-1}$ the damping of the free electrons gas.

\subsection{B. Power dissipated in a dipolar absorber}

Equation~(1) of the main text gives the power dissipated in a dipolar absorber plunged in a complex environment and illuminated by an incident field. The absorption is expressed as the difference between the extinction and the scattered power. It is a classical expression but we present its derivation here for the sake of completeness. In contrast to the standard demonstration that can be found in some textbooks, we provide a derivation that carefully handles the singularity of the field radiated by a dipole source on itself.

Energy conservation implies that the flux of the Poynting vector through a closed surface $\Sigma$ surrounding the absorber (and only the absorber) is negative and equal to the opposite of the power $P_a$ dissipated inside $\Sigma$,

\begin{equation}\label{eq:EqPa1}
    P_a = -\iint_{\Sigma} \frac{1}{2}\mathrm{Re}(\mathbf{E}^* \times \mathbf{H}) \cdot  d{\bf S} ,
\end{equation}

\noindent where $\mathbf{E}$ and $\mathbf{H}$ are the total electric and magnetic fields and $\mathbf{E}^*$ is the conjugate of $\mathbf{E}$. Usually, the total field is separated in two contributions, ${\bf E}({\bf r}) = {\bf E}_b({\bf r}) + {\bf E}_s({\bf r})$, where ${\bf E}_b$ is the field \emph{in the absence of} the absorber and ${\bf E}_s$ is the field scattered by the absorber located at ${\bf r} = {\bf r}_0$. By doing so, the dissipated power given by Eq.~\eqref{eq:EqPa1} can be expressed as the sum of two surface integrals. The first one corresponds to the extinction; it is given by $\iint_{\Sigma} \frac{1}{2}\mathrm{Re}(\mathbf{E}_s^* \!\times\! \mathbf{H}_b + \mathbf{E}_b^* \!\times\! \mathbf{H}_s) \!\cdot\!  d{\bf S}$. The second integral corresponds to the power scattered by the absorber in the presence of the environment; it is given by $\iint_{\Sigma} \frac{1}{2}\mathrm{Re}(\mathbf{E}_s^* \!\times\! \mathbf{H}_s) \!\cdot\!  d{\bf S}$. The latter corresponds to the power emitted by the induced dipole ${\bf p}$ since ${\bf E}_s({\bf r}) = \mu_0\omega^2 \mathbf{G}({\bf r},{\bf r}_0)\mathbf{p}$, with $\mathbf{G}({\bf r},{\bf r}_0)$ the Green tensor of the environment. This power can be expressed as a function of the imaginary part of $\mathbf{G}({\bf r}_0,{\bf r}_0)$. However, such derivation is not straightforward if one wants to properly handle the singularity of the real part of ${\bf G}({\bf r}_0,{\bf r}_0)$.

\medskip

In order to avoid the mathematical difficulty related to the Green tensor singularity, we present here a derivation that is based on a different decomposition of the total field. We write the total field as ${\bf E}({\bf r}) = {\bf E}_0({\bf r}) + {\bf E}_r({\bf r})$, where ${\bf E}_0({\bf r}) = \mu_0\omega^2 \mathbf{G}_0({\bf r},{\bf r}_0)\mathbf{p}$ is the field radiated by the induced dipole ${\bf p}$ \emph{in an homogeneous medium of refractive index $n$} and ${\bf E}_r({\bf r}) = \mathbf{E}_b({\bf r}) + \mu_0 \omega^2\mathbf{S}({\bf r},{\bf r}_0)\mathbf{p}$, with ${\bf S} = {\bf G} - {\bf G}_0$. The field ${\bf E}_0({\bf r})$ is singular at ${\bf r} = {\bf r}_0$, while the field ${\bf E}_r({\bf r})$ is regular everywhere in $\Sigma$. The dissipated power can be written as

\begin{align}\label{eq:EqPa2}
    P_a & = -\iint_{\Sigma} \frac{1}{2}\mathrm{Re}\big[(\mathbf{E}_r^*+\mathbf{E}_0^*) \times (\mathbf{H}_r+\mathbf{H}_0)\big] \cdot d{\bf S} \\
     & = -\iint_{\Sigma} \frac{1}{2} \Big[ \mathrm{Re}(\mathbf{E}_r^* \times \mathbf{H}_r) + \mathrm{Re}(\mathbf{E}_0^* \times \mathbf{H}_0) + \mathrm{Re}(\mathbf{E}_r^* \times \mathbf{H}_0 + \mathbf{E}_0^* \times \mathbf{H}_r) \Big] \cdot d{\bf S} \,.
\end{align}

The dissipated power is the sum of three different terms. Let us have a look at the first two terms:

\begin{itemize}
    \item $\iint_{\Sigma} \frac{1}{2}\mathrm{Re}(\mathbf{E}_r^* \times \mathbf{H}_r) \cdot d{\bf S} = 0$ since $\mathbf{E}_r$ and $\mathbf{H}_r$ verify Maxwell's equations without source inside $\Sigma$.

    \item $\iint_{\Sigma} \frac{1}{2}\mathrm{Re}(\mathbf{E}_0^* \times \mathbf{H}_0) \cdot d{\bf S}$ is the power emitted by the dipole source ${\bf p}$ in a homogeneous medium of refractive index $n$. It is given by the Larmor formula $\iint_{\Sigma} \frac{1}{2}\mathrm{Re}(\mathbf{E}_0^* \times \mathbf{H}_0) \cdot d{\bf S} = \frac{\omega^4n}{12\pi\varepsilon_0c^3}|\mathbf{p}|^2$. Since the imaginary part of the bulk Green tensor is given by $\mathrm{Im}[{\bf G}_0({\bf r}_0,{\bf r}_0)] = n\frac{\omega}{6\pi c}{\bf I}$, we can write $ \iint_{\Sigma} \frac{1}{2}\mathrm{Re}(\mathbf{E}_0^* \times \mathbf{H}_0) \cdot d{\bf S}= \frac{1}{2}\mu_0\omega^3 \mathbf{p}^\dagger\mathrm{Im}[{\bf G}_0({\bf r}_0,{\bf r}_0)]\mathbf{p}$.
\end{itemize}

\medskip

Now let us consider the third term in Eq.~(4). We use the Lorentz reciprocity formula, which relates two time-harmonic solutions of Maxwell's equations $(\mathbf{E}_1, \mathbf{H}_1,\omega_1,{\bf j}_1)$ and $(\mathbf{E}_2, \mathbf{H}_2,\omega_2,{\bf j}_2)$, where $(\mathbf{E}_i, \mathbf{H}_i)$ is the electromagnetic field created by the current distribution ${\bf j}_i$ at the frequency $\omega_i$. A general form of Lorentz reciprocity formula and its derivation can be found for instance in Annex 3 of Ref.~\cite{Lalanne2018}. Here we consider two solutions of Maxwell's equations inside the closed surface $\Sigma$ at the same frequency $\omega_1 = \omega_2 = \omega$. As the first solution, we consider the field $(\mathbf{E}_0, \mathbf{H}_0)$ radiated by the dipole ${\bf p}$, i.e., a current distribution ${\bf j}_1 = -i\omega \mathbf{p}\delta({\bf r} - {\bf r}_0)$. As the second solution, we consider the regular field $(\mathbf{E}_r^* , -\mathbf{H}_r^*)$ that is solution to Maxwell's equations without source, i.e., a current distribution  ${\bf j}_2 = 0$. We obtain (see Annex 3 of Ref.~\cite{Lalanne2018})

\begin{equation}
    \iint_{\Sigma} \Big[ (\mathbf{E}_r^* \times \mathbf{H}_0 + \mathbf{E}_0 \times \mathbf{H}_r^*) \Big] \cdot d{\bf S} = i\omega \mathbf{E}_r({\bf r}_0)^\dagger\mathbf{p} ,
\end{equation}

\noindent where the superscript $^\dagger$ denotes the conjugate transpose. One can then deduce the real part of this expression that takes the following form

\begin{equation}
    \iint_{\Sigma}\mathrm{Re}\big(\mathbf{E}_r^* \times \mathbf{H}_0 + \mathbf{E}_0^* \times \mathbf{H}_r \big) \cdot d{\bf S} = -\omega \mathrm{Im}[\mathbf{E}_r({\bf r}_0)^\dagger\mathbf{p}] = \omega \mathrm{Im}[\mathbf{p}^\dagger\mathbf{E}_r({\bf r}_0)]
\end{equation}

\bigskip

Finally, the power dissipated in the dipolar absorber becomes

\begin{equation}\label{eq:EqPa3}
    P_a = -\frac{1}{2}\omega \mathrm{Im}[\mathbf{p}^\dagger\mathbf{E}_r({\bf r}_0)] - \frac{1}{2}\mu_0\omega^3 \mathbf{p}^\dagger\mathrm{Im}[{\bf G}_0({\bf r}_0,{\bf r}_0)]\mathbf{p} ,
\end{equation}

\noindent with ${\bf E}_r({\bf r}_0) = \mathbf{E}_b({\bf r}_0) + \mu_0 \omega^2\mathbf{S}({\bf r}_0,{\bf r}_0) \mathbf{p}$ and $\mathbf{S} = \mathbf{G} - \mathbf{G}_0$. Thus, it follows

\begin{equation}\label{eq:EqPa4}
    P_a = -\frac{1}{2}\omega \mathrm{Im}[\mathbf{p}^\dagger\mathbf{E}_b({\bf r}_0)] - \frac{1}{2}\mu_0\omega^3 \mathrm{Im}[\mathbf{p}^\dagger\mathbf{S}({\bf r}_0,{\bf r}_0)\mathbf{p}] - \frac{1}{2}\mu_0\omega^3 \mathbf{p}^\dagger\mathrm{Im}[{\bf G}_0({\bf r}_0,{\bf r}_0)]\mathbf{p} .
\end{equation}

\noindent By using the relation $\mathrm{Im}(\mathbf{p}^\dagger\mathbf{S}\mathbf{p}) = \frac{1}{2i}\big(\mathbf{p}^\dagger\mathbf{Sp}-\mathbf{p}^\dagger\mathbf{S}^\dagger\mathbf{p}\big) = \mathbf{p}^\dagger \frac{\mathbf{S}-\mathbf{S}^\dagger}{2i} \mathbf{p}$, we can write the dissipated power as

\begin{equation}\label{eq:EqPa5}
    P_a = -\frac{1}{2}\omega \mathrm{Im}[\mathbf{p}^\dagger\mathbf{E}_b({\bf r}_0)] - \omega \mathbf{p}^\dagger\mathbf{s}\mathbf{p} - \frac{1}{2}\mu_0\omega^3 \mathbf{p}^\dagger\mathrm{Im}[{\bf G}_0({\bf r}_0,{\bf r}_0)]\mathbf{p} ,
\end{equation}

\noindent with ${\bf s} = \frac{1}{2}\mu_0\omega^2\frac{\mathbf{S}({\bf r}_0,{\bf r}_0)-\mathbf{S}({\bf r}_0,{\bf r}_0)^\dagger}{2i}$. \\Therefore, by defining the matrix ${\bf g}$ as $\mathbf{g} = {\bf s} + \frac{1}{2}\mu_0\omega^2\mathrm{Im}[{\bf G}_0({\bf r}_0,{\bf r}_0)] = {\bf s} + g_0{\bf I}$, we can express the power dissipated by the dipolar absorber as [see Eq.~(1) in the main text]

\begin{equation}\label{eq:PabsSI}
    P_a = - \frac{1}{2}\omega \mathrm{Im}[{\bf p}^\dagger {\bf E}_b({\bf r}_0)] - \omega {\bf p}^\dagger {\bf g}{\bf p}  \,.
\end{equation}

\noindent Note that, in the case where the materials composing the environment are reciprocal, the tensors ${\bf G}$ and ${\bf S}$ are symmetric. Thus, the expressions of ${\bf g}$ and ${\bf s}$ can be simplified as ${\bf s} = \frac{1}{2}\mu_0\omega^2\frac{\mathbf{S}({\bf r}_0,{\bf r}_0)-\mathbf{S}({\bf r}_0,{\bf r}_0)^*}{2i} = \frac{1}{2}\mu_0\omega^2 \mathrm{Im}[\mathbf{S}({\bf r}_0,{\bf r}_0)]$ and ${\bf g} = \frac{1}{2}\mu_0\omega^2 \mathrm{Im}[\mathbf{G}({\bf r}_0,{\bf r}_0)]$. We emphasize that the equations presented in the main text are valid for non-reciprocal  materials, provided that ${\bf g}$ is properly defined.

\subsection{C. Detailed derivation of the upper bound $P_a^\mathrm{max}$}

The power dissipated in a dipolar absorber is given by Eq.~(\ref{eq:PabsSI}) [Eq.~(1) in the main text]. In order to derive an upper bound that is independent of the absorber, we calculate the maximum of $P_a = f({\bf p})$.

We have seen in Section B that, in a complex medium composed of passive materials, the quantity $\omega {\bf u}^\dagger{\bf g}{\bf u}$ gives the total electromagnetic power emitted by the dipole source ${\bf u}$. Therefore, ${\bf u}^\dagger{\bf g}{\bf u} \geqslant 0$ whatever the vector ${\bf u}$ and ${\bf g}$ is a semi-definite positive matrix. Let us use this important property to derive the maximum of $P_a = f({\bf p})$. For that purpose, we have to rewrite the dissipated power under the form $P_a = -\omega {\bf u}^\dagger{\bf g}{\bf u} + A$, where $A$ is independent of ${\bf p}$.

This can be done with the following algebraic manipulations

\begin{align}
  P_a &= - \omega {\bf p}^\dagger {\bf g}{\bf p} - \frac{1}{2}\omega \mathrm{Im}({\bf p}^\dagger {\bf E}_b)
  = - \omega {\bf p}^\dagger {\bf g}{\bf p} \,+\, \frac{1}{4i}\omega \left ( {\bf E}_b^\dagger{\bf p} - {\bf p}^\dagger {\bf E}_b \right )\notag\\
  &= - \omega {\bf p}^\dagger {\bf g}{\bf p} \,+\, \frac{1}{4i}\omega \left ( {\bf E}_b^\dagger{\bf g}^{-1}{\bf g}{\bf p} - {\bf p}^\dagger{\bf g}{\bf g}^{-1}{\bf E}_b \right ) \notag\\
  &= -\omega \left ( {\bf p}^\dagger {\bf g}{\bf p} + \frac{i}{4}{\bf E}_b^\dagger{\bf g}^{-1}{\bf g}{\bf p} - \frac{i}{4}{\bf p}^\dagger{\bf g}{\bf g}^{-1}{\bf E}_b  \right ) \notag\\
  &= -\omega \left [ \left ( {\bf p}^\dagger + \frac{i}{4}{\bf E}_b^\dagger{\bf g}^{-1} \right ) {\bf g}{\bf p} - {\bf p}^\dagger{\bf g} \frac{i}{4}{\bf g}^{-1}{\bf E}_b  - \frac{i}{4}{\bf E}_b^\dagger{\bf g}^{-1}{\bf g}\frac{i}{4}{\bf g}^{-1}{\bf E}_b
  + \frac{i}{4}{\bf E}_b^\dagger{\bf g}^{-1}{\bf g}\frac{i}{4}{\bf g}^{-1}{\bf E}_b \right ] \notag\\
  &= -\omega \left [ \left ( {\bf p}^\dagger + \frac{i}{4}{\bf E}_b^\dagger{\bf g}^{-1} \right ) {\bf g}{\bf p} - \left ( {\bf p}^\dagger + \frac{i}{4}{\bf E}_b^\dagger{\bf g}^{-1} \right ) {\bf g} \frac{i}{4}{\bf g}^{-1}{\bf E}_b - \frac{1}{16}{\bf E}_b^\dagger{\bf g}^{-1}{\bf E}_b \right ] \notag\\
  &= -\omega \left ( {\bf p}^\dagger + \frac{i}{4}{\bf E}_b^\dagger{\bf g}^{-1} \right ) {\bf g} \left ( {\bf p} - \frac{i}{4}{\bf g}^{-1}{\bf E}_b \right ) + \frac{\omega}{16}{\bf E}_b^\dagger{\bf g}^{-1}{\bf E}_b \,.
\end{align}

\noindent Now the dissipated power takes the form $P_a = -\omega {\bf u}^\dagger{\bf g}{\bf u} + A$, with ${\bf u} = {\bf p} - \frac{i}{4}{\bf g}^{-1}{\bf E}_b$ and $A = \frac{\omega}{16}{\bf E}_b^\dagger{\bf g}^{-1}{\bf E}_b$. This specific form of the dissipated power in a dipolar absorber corresponds to Eq.~(2) in the main text. Since ${\bf g}$ is a semi-definite positive matrix, we know that $-\omega ( {\bf p}^\dagger + \frac{i}{4}{\bf E}_b^\dagger{\bf g}^{-1} ) {\bf g} ( {\bf p} - \frac{i}{4}{\bf g}^{-1}{\bf E}_b ) \leqslant 0$. This readily leads to the upper bound derived in the main text, which is independent of the absorber and depends only on the environment and the illumination,

\begin{equation}\label{eq:EqPmaxSI}
    P_a^\mathrm{max} = \frac{\omega}{16}\mathbf{E}_b^\dagger\mathbf{g}^{-1}\mathbf{E}_b \,.
\end{equation}

\subsection{D. Detailed derivation of the optimal polarizability $\alpha_\mathrm{opt}$}

We provide in this Section more details on the derivation of the optimal polarizability $\boldsymbol{\alpha}_\mathrm{opt}$ given by Eq.~(4) in the main text. According to Eq.~\eqref{eq:EqPmaxSI}, the maximum absorption is reached for an optimal dipole

\begin{equation}\label{eq:pmax}
    {\bf p}_\mathrm{opt} = \frac{i}{4}\mathbf{g}^{-1}\mathbf{E}_b \,.
\end{equation}

\noindent The induced dipole, located at ${\bf r} = {\bf r}_0$, is given by

\begin{equation}\label{eq:pind}
    {\bf p} = \varepsilon_0 \boldsymbol{\alpha}(\omega) \left [ \mathbf{E}_b + \mu_0 \omega^2\mathbf{S}({\bf r}_0,{\bf r}_0) \mathbf{p} \right ],
\end{equation}

\noindent where $\boldsymbol{\alpha}(\omega)$ is the polarizability tensor of the dipolar absorbing particle and $\mathbf{S} = \mathbf{G} - \mathbf{G}_0$, with $\mathbf{G}$ the Green tensor of the environment (without absorber) and $\mathbf{G}_0$ the Green tensor of the homogeneous medium of refractive index $n$ that surrounds the particle.

By injecting Eq.~\eqref{eq:pmax} into Eq.~\eqref{eq:pind}, one finds that, in order to reach the upper bound, the absorber polarizability tensor has to fulfill the expression

\begin{equation}\label{eq:palpha}
    \frac{i}{4}\mathbf{g}^{-1} = \varepsilon_0 \boldsymbol{\alpha}_\mathrm{opt}(\omega) \left [ \mathbf{I} + \mu_0 \omega^2\mathbf{S}({\bf r}_0,{\bf r}_0) \frac{i}{4}\mathbf{g}^{-1} \right ] .
\end{equation}

\noindent Right multiplying both sides of the equation by $-4i{\bf g}$ and isolating the polarizability leads to

\begin{equation}\label{eq:versEqM2}
    \boldsymbol{\alpha}_\mathrm{opt}(\omega) = \mu_0 c^2 \left [ \mu_0\omega^2\mathbf{S}({\bf r}_0,{\bf r}_0) -4i\mathbf{g} \right ]^{-1} .
\end{equation}

\noindent Since, in the case of reciprocal materials, $\mathbf{g} = \frac{1}{2}\mu_0\omega^2 \mathrm{Im}[{\bf G}({\bf r}_0,{\bf r}_0)] = \frac{1}{2}\mu_0\omega^2 \mathrm{Im}[{\bf G}_0({\bf r}_0,{\bf r}_0)] + \frac{1}{2}\mu_0\omega^2 \mathrm{Im}[{\bf S}({\bf r}_0,{\bf r}_0)] = \frac{\omega^3n}{12\pi\varepsilon_0c^3}{\bf I} + \frac{1}{2}\mu_0\omega^2 \mathrm{Im}[{\bf S}({\bf r}_0,{\bf r}_0)]$, we have the following relation

\begin{align*}\label{eq:versEqM2bis}
 \mu_0\omega^2\mathbf{S}({\bf r}_0,{\bf r}_0) - 4i\mathbf{g} = & \mu_0\omega^2\mathbf{S}({\bf r}_0,{\bf r}_0) - 2i\mu_0\omega^2\mathrm{Im}[\mathbf{S}({\bf r}_0,{\bf r}_0)] - i\frac{\omega^3n}{3\pi\varepsilon_0c^3}{\bf I}  \\
 = & \mu_0\omega^2\mathbf{S}^*({\bf r}_0,{\bf r}_0) - i\frac{\omega^3n}{3\pi\varepsilon_0c^3}{\bf I} \,,
\end{align*}

\noindent with $\mathbf{S}^*$ the conjugate of $\mathbf{S}$. Therefore, the optimal polarizability given by Eq.~\eqref{eq:versEqM2} can be rewritten as

\begin{equation}
    \boldsymbol{\alpha}_\mathrm{opt}(\omega) = \mu_0 c^2 \left [ \mu_0\omega^2\mathbf{S}^*({\bf r}_0,{\bf r}_0) - i\frac{\omega^3n}{3\pi\varepsilon_0c^3}{\bf I} \right ]^{-1} ,
\end{equation}

\noindent which is equivalent to Eq.~(4) in the main paper,

\begin{equation}\label{eq:EqM2SI}
    \boldsymbol{\alpha}_\mathrm{opt}(\omega) = \frac{c^2}{\omega^2}\left [ \mathbf{S}^*({\bf r}_0,{\bf r}_0)- i\frac{\omega n}{3\pi c} \mathbf{I} \right ]^{-1} .
\end{equation}

\noindent Note that, in the case of non-reciprocal materials, the conjugate $\mathbf{S}^*$ has simply to be replaced by the conjugate transpose $\mathbf{S}^\dagger$.

\subsection{E. Link with the antenna gain in the scalar approximation and under plane-wave illumination}

It is known from antenna theory that, when considering an antenna receiving a signal from the direction $(\theta,\phi)$, the maximum absorption cross-section of the load is given by $G_a(\theta,\phi)\lambda^2/(4\pi n^2)$, with $G_a(\theta,\phi)$ the antenna gain~\cite{Balanis}. This upper bound is reached if the load is impedance-matched with the antenna~\cite{Balanis}. We demonstrate hereafter that this simple expression is valid for an optical nanoantenna illuminated by a plane wave, \emph{provided that the nanoantenna can be described in the scalar approximation}.

\medskip

The expression of the absorption upper bound under plane-wave illumination and in the scalar approximation is given by Eq.~(7) in the main text,

\begin{equation}\label{eq:EqM1scalarSI}
    \sigma_a^\mathrm{max} \approx \frac{3\lambda^2}{8\pi n^2} \frac{|E_{bz}|^2}{|E_\mathrm{inc}|^2} \frac{g_0}{g_{zz}} \,.
\end{equation}

\noindent Note that, by scalar approximation, we mean that one component of the electromagnetic field (here the $z$ component) is dominant over the two others. Our objective is to introduce the antenna gain $G_a$ in this expression.

\medskip

Let us first discuss the antenna in the emitting mode in order to define its gain and its directivity. These definitions are usual but we recall them for the sake of completeness. We consider an electric dipole linearly polarized along the dominant direction, the $z$ direction, and located close to the nanoantenna at the position ${\bf r} = {\bf r}_0$ of the absorber, ${\bf p}_e = p_e{\bf u}_z$ with ${\bf u}_z$ the unitary vector of the $z$ direction. We define the gain $G_a(\theta,\phi)$ of the antenna as the ratio between the power $P_r(\theta,\phi)$ radiated by this dipole in the direction $(\theta,\phi)$ and the total power $P_T$ emitted by this dipole,

\begin{equation}\label{eq:defGain}
    G_a(\theta,\phi) = 4\pi\frac{P_r(\theta,\phi)}{P_T} .
\end{equation}

\noindent The antenna directivity $D(\theta,\phi)$ is defined as

\begin{equation}
    D(\theta,\phi) = 4\pi\frac{P_r(\theta,\phi)}{P_r} ,
\end{equation}

\noindent where $P_r$ is the power radiated in all directions, $P_r = \iint P_r(\theta,\phi)\sin(\theta) d\theta d\phi$. Since $P_T = P_{nr} + P_r$, with $P_{nr}$ the non-radiative power (i.e., the power absorbed in the antenna in emitting mode), gain and directivity are related by $G_a(\theta,\phi) = \eta_r D(\theta,\phi)$, with $\eta_r = P_r/P_T$ the radiative efficiency of the antenna.

\medskip

We come back to the problem of the antenna in the receiving mode. To express the upper bound of the absorption cross-section of the absorber (the load) as a function of the antenna gain, we need to rewrite Eq.~\eqref{eq:EqM1scalarSI} as a function of $P_r(\theta,\phi)$ and $P_T$. Introducing the total power in Eq.~\eqref{eq:EqM1scalarSI} is straightforward since, within the scalar approximation,

\begin{equation}\label{eq:PT}
    P_T = \omega{\bf p}_e^\dagger{\bf g}{\bf p}_e = \omega|p_e|^2g_{zz} .
\end{equation}

\begin{figure}[b]
	%\centerline{\includegraphics[width=0.7\columnwidth]{Figure/Figure1_SI.pdf}}
    \centerline{\includegraphics[width=0.7\columnwidth]{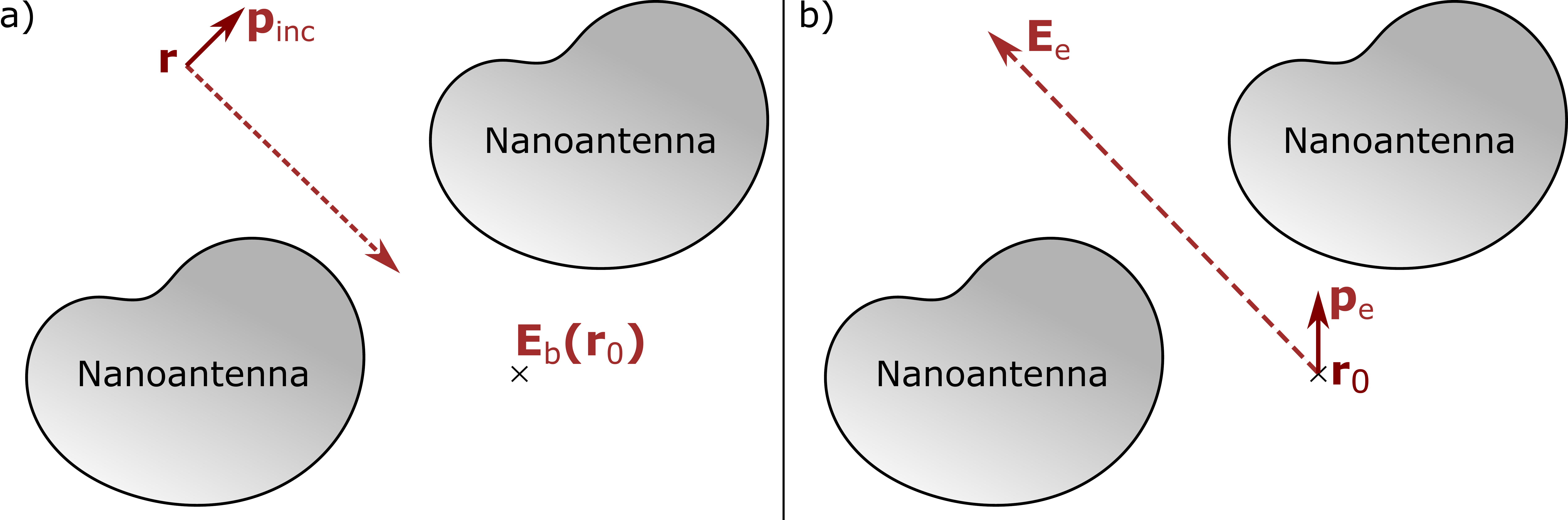}}
	\caption{(a) An incident plane wave ${\bf E}_\mathrm{inc}$ is created by an electric dipole ${\bf p}_\mathrm{inc}$ located in the far-field at a distance $\bf{r}$ along the direction $(\theta,\phi)$. It is incident on the antenna (without absorber) and creates an electric field ${\bf E}_b$ at ${\bf r}_0$. (b) An electric dipole ${\bf p}_e = p_e{\bf u}_z$ located at ${\bf r}_0$ radiates an electric field ${\bf E}_e$ in the far-field at a distance $\bf{r}$ along the direction $(\theta,\phi)$.}
	\label{fig:principeSI}
\end{figure}

\noindent To introduce the radiated power $P_r(\theta,\phi)$ in Eq.~\eqref{eq:EqM1scalarSI}, we need to make the link between the receiving mode and the emitting mode with the Lorentz reciprocity theorem. A first problem is the antenna in the receiving mode; a plane wave ${\bf E}_\mathrm{inc} = E_\mathrm{inc}{\bf u}_z$ is incident on the antenna (without absorber) from the direction $(\theta,\phi)$ and creates an electric field ${\bf E}_b$ at ${\bf r}_0$, see Fig.~1(a). A second problem is the antenna in the emitting mode; an electric dipole ${\bf p}_e = p_e{\bf u}_z$ located at ${\bf r}_0$ radiates an electric field ${\bf E}_e$ in the far-field at a distance $r$ along the direction $(\theta,\phi)$, see Fig.~1(b). To link these two problems by reciprocity, we first need to write that the incident plane wave of the first problem results from the emission of an electric dipole ${\bf p}_\mathrm{inc}$ located in the far-field at a distance $r$ along the direction $(\theta,\phi)$. This dipole is related to the incident field ${\bf E}_\mathrm{inc}$ through the relation

\begin{equation}\label{eq:pinc}
    {\bf p}_\mathrm{inc} = \frac{4\pi}{\mu_0\omega^2}\frac{r}{e^{ik_0nr}} {\bf E}_\mathrm{inc} .
\end{equation}

%We introduce the vector ${\bf u}$ that defines the polarization of the incident plane wave, ${\bf E}_\mathrm{inc} = E_\mathrm{inc}{\bf u}$, with ${\bf u}$ an unitary vector contained in the plane perpendicular to the direction $(\theta,\phi)$. Note that the vector ${\bf u}$ is real in the case of a linearly polarized plane wave but complex in the case of a circularly polarized plane wave, with a $\pi/2$ phase shift between the two components of ${\bf u}$ in the plane perpendicular to the direction $(\theta,\phi)$. The dipole ${\bf p}_\mathrm{inc}$ has the same polarization as the incident field ${\bf E}_\mathrm{inc}$.

\noindent Within the scalar approximation, Lorentz reciprocity between the receiving antenna and the emitting antenna can be expressed as \cite{Biagioni2012}

\begin{equation}
    p_e E_{bz} = p_\mathrm{inc} E_{ez} .
\end{equation}

\noindent Using Eq.~\eqref{eq:pinc} leads to an expression of the local intensity $|E_{bz}|^2$ in the receiving mode as a function of the far-field intensity $|E_{ez}|^2$ in the emitting mode

\begin{equation}\label{eq:recip}
    |E_{bz}|^2 = \frac{16\pi^2}{(\mu_0\omega^2)^2} r^2 \frac{|E_\mathrm{inc}|^2}{|p_e|^2} |E_{ez}|^2 .
\end{equation}

\noindent Moreover, the far-field intensity $|E_{ez}|^2$ in the emitting mode is related to the radiated power $P_r(\theta,\phi)$ through the expression

\begin{equation}\label{eq:Pr}
    P_r(\theta,\phi) = \frac{1}{2\mu_0 c}nr^2 |E_{ez}|^2 .
\end{equation}

\noindent Combining Eqs.~\eqref{eq:recip} and~\eqref{eq:Pr} leads to

\begin{equation}\label{eq:recip2}
    \frac{|E_{bz}|^2}{|E_\mathrm{inc}|^2} = \frac{32\pi^2c}{\mu_0\omega^4n|p_e|^2} P_r(\theta,\phi) .
\end{equation}

\noindent Finally, by injecting Eqs.~\eqref{eq:PT} and~\eqref{eq:recip2} into Eq~\eqref{eq:EqM1scalarSI} leads to the expression of the maximum absorption cross-section as a function of the antenna gain in the scalar approximation

\begin{equation}
    \sigma_a^\mathrm{max} = \frac{\lambda^2}{4\pi n^2} 4\pi \frac{P_r(\theta,\phi)}{P_T} = \frac{\lambda^2}{4\pi n^2} G_a(\theta,\phi)  \,.
\end{equation}

\noindent Note that we have used the expression $g_0 = \frac{\omega^3 n}{12\pi \varepsilon_0c^3}$.

\medskip

Let us emphasize once again that the simple relation between the absorption upper bound and the antenna gain is only valid for an antenna in the scalar approximation and illuminated by a plane wave. If the field is not dominated by a single component, it is not possible to express the maximum absorption cross-section as a function of the gain. One should define several gain values for different polarizations of the source and different polarizations of the emission.

\subsection{F. Reciprocal materials : decomposition of the vectorial problem in the sum of three scalar problems}

We consider in this Section the case where the environment is composed of reciprocal materials. In that case, we show that it is always possible to find an orthonormal basis where the vectorial problem can be written as the sum of three scalar problems.

Since $\mathbf{g}$ is a real and symmetric matrix, it can be diagonalized and its eigenvectors form an orthonormal basis, $\mathbf{g} = ^t\mathbf{P}\mathbf{D}\mathbf{P}$ where $\mathbf{D}$ is the diagonal matrix formed with the eigenvalues $d_i$ of $\mathbf{g}$ (which are real and positive) and $\mathbf{P}$ is the matrix formed by the eigenvectors ${\bf v}_i$, with $^t\mathbf{P} \mathbf{P} = \mathbf{P} ^t\mathbf{P} = \mathbf{I}$. The superscript $^t$ denotes matrix transposition.

Therefore, if we express the vector ${\bf E}_b$ and the matrix ${\bf g}$ in the eigenvectors basis $({\bf v}_1,{\bf v}_2,{\bf v}_3)$, the equations Eq.~(3), Eq.~(5) and Eq.~(6) of the main text take the form

\begin{equation}\label{eq:EqPmaxsca}
    P_a^\mathrm{max} = \frac{\omega}{16}\sum_{i=1}^3 \frac{|E_b^i|^2}{d_i} ,
\end{equation}

\begin{equation}\label{eq:EqEffmaxsca}
    \eta_a^\mathrm{max} = \frac{\omega}{16}\sum_{i=1}^3 \frac{|E_b^i|^2}{d_i P_\mathrm{inc}} ,
\end{equation}

\begin{equation}\label{eq:EqM1sca}
    \sigma_a^\mathrm{max} = \frac{3\lambda^2}{8\pi n^2} \sum_{i=1}^3 \frac{g_0}{d_i} \frac{|E_b^i|^2}{|{E}_\mathrm{inc}|^2} ,
\end{equation}

\noindent with $d_i$ the eigenvalues of ${\bf g}$ and $E_b^i = {\bf E_b}\cdot{\bf v}_i$.

\subsection{G. Comparison between the predicted upper bound and a rigorous numerical calculation}

We validate our theoretical predictions of the upper bound and the optimal polarizability by calculating rigorously the absorption cross-section of a nanosphere, which has the optimal polarizability given by Eq.~\eqref{eq:EqM2SI}. The validation is done here on the Yagi-Uda example but the same kind of verification has been done for the other examples. Rigorous calculations are performed with a multipole method (also known as T-matrix method), which expands the electromagnetic field onto the basis of vector spherical harmonics. We have used in the expansion a multipolar order $N=15$.

\begin{figure}
	%\centerline{\includegraphics[width=0.7\columnwidth]{Figure/Figure1_SI.pdf}}
    \centerline{\includegraphics[width=0.45\columnwidth]{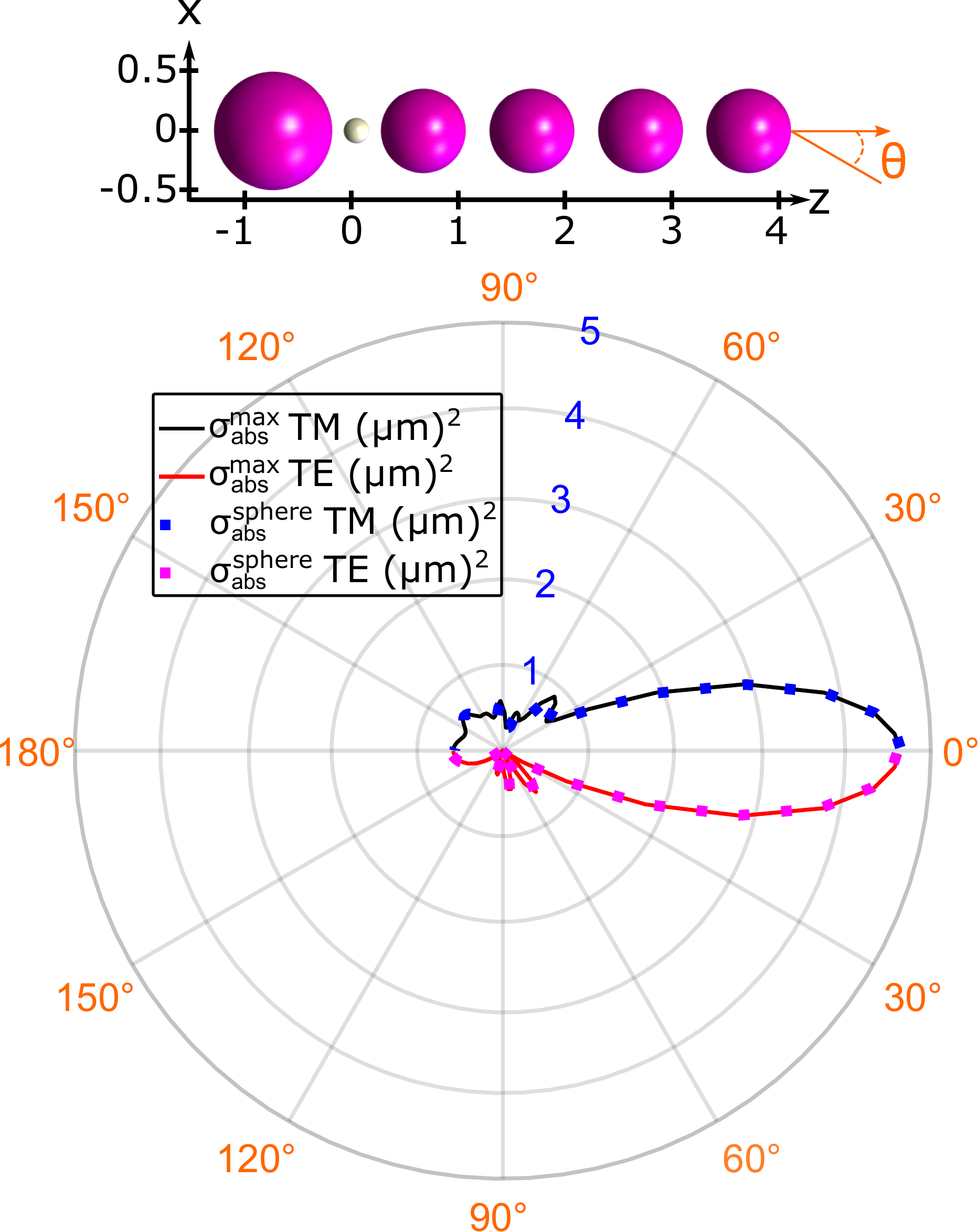}}
	\caption{Direct absorption cross-section calculation for an absorbing nanosphere (in white, radius 176~nm and permittivity $\varepsilon=-6.46+1.22i$) inserted inside the non-absorbing Yagi-Uda antenna (purple spheres, permittivity and radii given in the main text). The blue squares curve corresponds to TM polarization and the purple squares curve to TE polarization. These rigorous calculations are superimposed on the upper bound (no absorber, Yagi-Uda antenna alone) predicted with Eq.~(6) of the main text (black and red solid curves).}
	\label{fig:SphereYagi}
\end{figure}

As mentioned in the main text, in order to reach the optimal polarizability, we have chosen a sphere of radius 176~nm filled with a material of relative permittivity $\varepsilon_\mathrm{opt} = -6.46+1.22i$ ~\footnote{We have rigorously calculated the relation between the electric-dipole polarizability and the sphere radius and permittivity with Mie theory}. We position the absorbing sphere in between the reflector and the first director, see the small white sphere in the top panel of Fig.~\ref{fig:SphereYagi}. The polar plot in Fig.~\ref{fig:SphereYagi} shows the absorption cross-section of this nanosphere as a function of the incident angle in TM (blue squares) or TE (purple squares) polarization calculated directly with the rigorous multipole method. On the other hand, the upper bound predicted by Eq.~(6) in the main text is shown with black and red curves for TM and TE incident polarizations, respectively. The perfect agreement between rigorous calculations and theoretical predictions confirms (i) the validity of our theoretical derivations, (ii) the possibility to reach the upper bound, and (iii) the validity of the dipolar approximation for the chosen spherical absorber.

%\end{suppinfo}

\bibliography{Publi_FM_ACSPhot_v1}

\end{document}